\documentclass[times]{aastex631}

\newcommand{\ciao}{\texttt{Ciao}}
\newcommand{\acisextract}{\texttt{ACISExtract}}
\newcommand{\sherpa}{\texttt{Sherpa}}

\newcommand{\source}{J1621}

\usepackage{color}
\usepackage{comment}

\usepackage{verbatim}

\defcitealias{Zola2017}{Z+17}

\begin{document}
\defcitealias{Zola2017}{Z+17}

\title{The Variable X-ray Emission of the Cataclysmic Variable V1460 Her with a Rapidly Rotating White Dwarf}

\author[0000-0001-8473-5140]{Erik B. Monson}
\affiliation{Department of Astronomy and Astrophysics, The Pennsylvania State University, 525 Davey Lab, University Park, PA 16802, USA}

\author[0000-0002-7481-5259]{George G. Pavlov}
\affiliation{Department of Astronomy and Astrophysics, The Pennsylvania State University, 525 Davey Lab, University Park, PA 16802, USA}

\author[0000-0002-7371-5416]{Gordon P. Garmire}
\affiliation{Huntingdon Institute for X-ray Astronomy, LLC, 10677 Franks Rd, Huntingdon, PA 16652, USA }

\begin{abstract}

Using a recent Chandra ACIS observation  of the eclipsing cataclysmic variable (CV) V1460 Her (2MASS J16211735+4412541) with a fast-rotating ($P_{\rm spin} = 38.9$~s) white dwarf, we estimated a flux $F_{0.5-7\, {\rm keV}}^{\rm abs} = (1.9\pm 0.5)\times 10^{-14}$ erg cm$^{-2}$ s$^{-1}$, a factor of $\sim7$ lower than found from previous Swift XRT observations $\approx8$ years ago, when the CV was quiescent.
The drop in the flux suggests a corresponding drop in the accretion rate, and the resulting intrinsic luminosity $L_{0.5-7~{\rm keV}} \sim 1.6\times 10^{29}~{\rm erg~s^{-1}}$ places V1460 Her among the lowest-luminosity magnetic CVs known, in a state of very low accretion.
\end{abstract}



\section{Introduction} \label{sec:intro}

The eclipsing cataclysmic variable (CV) V1460 Her (2MASS J16211735+4412541; \source\ hereafter) consists of a white dwarf (WD) primary 
and a K-type secondary, with a binary period $P_{\rm bin}=4.99$ hours \citep{Lohr2013}. 
The system has been observed to produce UV \citep{Ashley2020} and optical $g$-band \citep{Pelisoli2021} pulsations, implying a WD spin period $P_{\rm spin}=38.9$ s \citep{Ashley2020}, the fourth fastest-rotating confirmed WD known. The rapid rotation suggests that J1621 is a candidate for a ``magnetic propeller,'' like the well-studied AE Aquarii, which exhibits complex X-ray variability \citep{Kitaguchi2014}.

\source\ was observed with Swift XRT on 2016 June 10 and 14 following an outburst on 2016 June 4, and on 2017 January 10, 13, 14 and 19 for 5.7 ks in total. \citet{Zola2017} (Z+17 hereafter) extracted a spectrum from the merged observations in a $47\farcs2$ radius aperture centered on \source. Assuming a power-law (PL) spectrum and a Galactic Hydrogen column density $N_H=1.6\times 10^{21}$ cm$^{-2}$ along the line of sight to the source, they estimated an absorption-corrected flux $F^{\rm unabs}_{0.3-10 \rm keV} = 2.2^{+5.0}_{-0.8}\times10^{-13}$ erg cm$^{-2}$ s$^{-1}$ and a photon index $\Gamma=1.9\pm1.2$. They interpreted the X-ray emission up to $\approx 6.5$ months after outburst as evidence of persistent accretion.

We re-reduced the XRT data using the \texttt{xrtpipeline} and \texttt{xrtproducts} scripts in \texttt{HEASoft}\footnote{\url{https://heasarc.gsfc.nasa.gov/docs/software/heasoft/}} v6.33.2. We extracted spectra in an $r = 28\farcs2$ aperture, and merged them following \citetalias{Zola2017}, finding 20 counts in the 0.5$-$7 keV band. We fit a PL model
with \sherpa\ v4.16, assuming $N_H = 1 \times 10^{20}~{\rm cm^{-2}}$, corresponding to the $E(B-V)=0.02\pm0.02$ quoted by \citet{Ashley2020}, and found 
$\Gamma = 2.0 \pm 0.4$ and $F_{\rm 0.5-7\,keV}^{\rm abs} = 1.6^{+0.5}_{-0.4}\times 10^{-13}$ erg cm$^{-2}$ s$^{-1}$, 
which virtually coincides with the unabsorbed flux at such low $N_H$. We also fit an \texttt{APEC} model (emission from an optically thin thermal plasma), finding 
$kT = 3.3^{+1.9}_{-1.5}$ keV 
and 
$F_{\rm 0.5-7\,keV}^{\rm abs} = (1.4\pm 0.4)\times 10^{-13}$ erg cm$^{-2}$ s$^{-1}$.


\section{Chandra Observation} \label{sec:obs}
To measure the spectrum of J1621 and look for flux variation with the binary rotation phase, it was 
observed with the Chandra ACIS detector (ObsID 27350, PI: Garmire; imaged on the S3 chip) on 2024 June 27 for 18 ks (one binary period). Following a standard data reduction with \ciao\ v4.16 and CALDB v4.11.2, we extracted photometry and spectra with \acisextract\ v2023aug14\footnote{\url{https://sites.psu.edu/acisextractandtarasoftware/acis-extract-software/}}. We detected 19 source counts (rather than the $\sim 500$ expected from the 2016-2017 flux), and a net count rate of $1.1\pm 0.2$ counts ks$^{-1}$ in the 90\% enclosed energy aperture (approximately circular, with $r\approx0\farcs85$ for this on-axis source). 

We fit the ACIS spectrum with PL and \texttt{APEC} models, assuming $N_H = 1 \times 10^{20}~{\rm cm^{-2}}$, and found a considerable flux drop from 2016-2017:
$F_{\rm 0.5-7\,keV}^{\rm abs} = (1.9\pm 0.5)\times 10^{-14}$ 
and 
$2.1^{+0.6}_{-0.5} \times 10^{-14}$ erg cm$^{-2}$ s$^{-1}$ 
for the PL and \texttt{APEC} fits, respectively,  $\approx 7$ times lower than the flux from our XRT re-analysis. 

\begin{figure}
\centering
\includegraphics[width=1.0\textwidth]{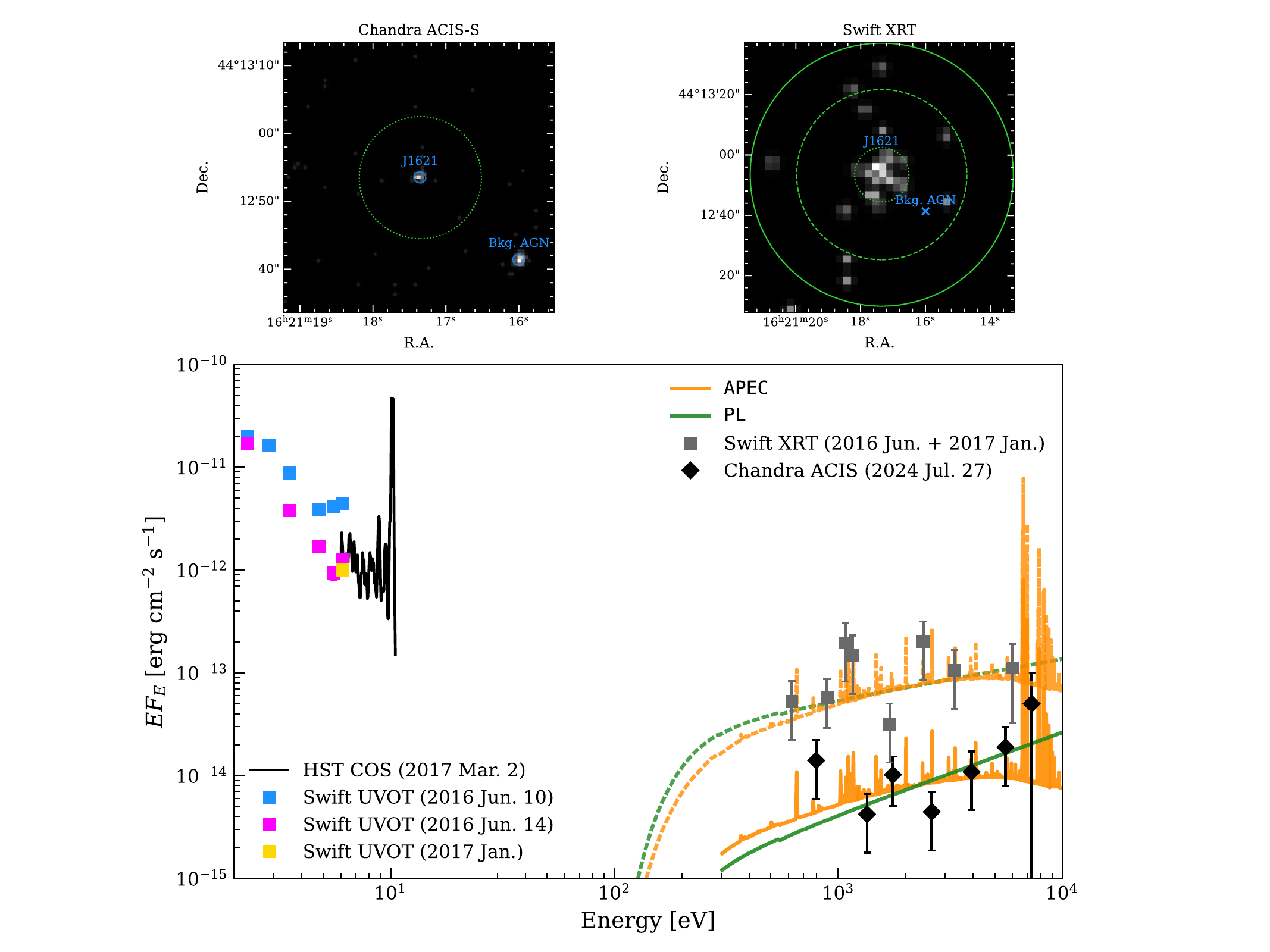}
\caption{\label{fig:spec}
\textit{Upper Left:} 
The 0.5--7 keV Chandra ACIS-S3 $40''\times 40''$ image of \source\ and its vicinity.
The \acisextract\ extraction regions are shown as blue polygons. 
\textit{Upper Right:} The co-added 2016 June + 2017 January Swift XRT $90''\times 90''$ images, with the $r=42\farcs7$ extraction aperture used by \citetalias{Zola2017} shown as a solid green circle. The dashed green circle with $r=28\farcs2$ is the extraction aperture for our re-analysis.
The position of the AGN is marked with a blue cross.In both upper panels the dotted $18''$ diameter green circle represents the approximate on-axis FWHM of the XRT PSF.
\textit{Lower Panel:} The ACIS spectrum (black diamonds), and the co-added XRT spectrum (gray squares), which may include a small contribution from the AGN (see text). For each spectrum, lines show the best-fitting PL (green) and \texttt{APEC} models (orange). For both X-ray observations, we used the best-fitting PL model to unfold the data from the response functions. We also show Swift UVOT measurements (colored squares) obtained throughout 2016-2017 \citepalias{Zola2017}, along with a boxcar-smoothed HST COS spectrum (black lines; the highest peak is the geocoronal Ly-$\alpha$ line) obtained in 2017 March \citep{Ashley2020}.}
\end{figure}

We find a photon index
$\Gamma = 1.3\pm 0.5$, 
consistent with the previous measurement given the large uncertainties. The \texttt{APEC} temperature is not constrained, with a best-fitting $kT\sim 6$ keV. We show the Chandra spectrum and the best-fitting models in the bottom panel of \autoref{fig:spec}, along with the archival Swift XRT and UVOT data and an HST COS UV spectrum\footnote{Available on MAST: \dataset[10.17909/5exd-xk84]{http://dx.doi.org/10.17909/5exd-xk84}}. 

In the ACIS image a second source is detected (27 counts), offset from \source\ by $\approx 19''$ (\autoref{fig:spec}, upper left panel). We identify this source with SDSS J162115.99+441241.2, a galaxy with SDSS photometric redshift $z = 0.216 \pm 0.041$\footnote{Retrieved from \url{https://skyserver.sdss.org/dr17}}. No SDSS spectrum is available. A spectral fit with a PL model, assuming $N_H = 1 \times 10^{20}~{\rm cm^{-2}}$ and $z = 0.216$, yields a 0.5-7 keV flux $3.0^{+0.7}_{-0.6}\times 10^{-14}$ erg cm$^{-2}$ s$^{-1}$ and an intrinsic luminosity $L_{\rm 2-10\,keV} = 3.9\times 10^{42}$ erg s$^{-1}$, consistent with an AGN. This AGN candidate is not discernible in the 2016-2017 XRT and UVOT
observations, though its position falls within the aperture used by 
\citetalias{Zola2017} to extract the XRT spectrum of J1621. If we assume that the AGN flux was the same in 2016-2017 as in
2024, we find that the AGN-subtracted XRT flux from J1621 is 
$1.1^{+0.9}_{-0.7} \times 10^{-13}$ erg cm$^{-2}$ s$^{-1}$, 
consistent with the measurement from \citetalias{Zola2017}.

\section{Discussion} \label{sec:disc}

Based on the Chandra PL fit and the distance $d=263.2\pm 1.4$ (inferred from Gaia EDR3; \citealp{gaiaEDR3}), 
we estimate the intrinsic luminosity of \source\ as 
$L_{\rm 0.5-7\,keV} \approx 1.7 \times 10^{29}~{\rm erg~s^{-1}}$
whereas the AGN-subtracted 2016-2017 XRT flux corresponds to
$L_{\rm 0.5-7\,keV}\approx 1.0\times 10^{30}~{\rm erg~s^{-1}}$.

Assuming accretion onto the WD surface, the accretion-powered luminosity of the WD can be estimated as
\begin{equation}
    L_{\rm accr} \sim G M_{\rm WD} \dot m / R_{\rm WD}.
\end{equation} 
If we take $M_{\rm WD} = 0.87 M_{\odot}$ \citep{Ashley2020}, $R_{\rm WD} = 7000$ km, 
the accretion rate during 2016-2017 was 
$\dot m \approx 6.1\times10^{12}\eta^{-1}~{\rm g~s^{-1}}$, 
dropping to 
$1.0\times10^{12}\eta^{-1}~{\rm g~s^{-1}}$ 
by 2024 June ($\eta = L_{\rm 0.5-7\,keV}/L_{\rm accr}$ is an unknown bolometric correction). 

Assuming the mass and radius above, we estimate the co-rotation radius as $r_c = 1.6\times 10^{9}\,{\rm cm}$ and the magnetospheric radius as $r_m \sim 1.0\times 10^{11} B_6^{4/7} \dot{m}_{12}^{-2/7}\,{\rm cm}\,,$ where $B=10^6 B_6$~G is the magnetic field, and $\dot{m} = 10^{12} \dot{m}_{12}$~g~s$^{-1}$ is the accretion rate \citep[see][equations 4 and 5]{Zavlin2004}. The criterion $r_c \geq r_m$ for the accretion flow to not be disrupted by the magnetic propeller effect yields $B\lesssim 750 \dot{m}_{12}^{1/2}$~G, implying an extremely weak magnetic field with $B\sim 10^2-10^3$~G for sustained accretion. A magnetic moment $\mu \sim 10^{30}~\rm G~cm^{3}$ similar to AE Aqr \citep{Patterson1994} places J1621 in the propeller state, with magnetospheric radius 2-8 times the corotation radius. Further observations remain necessary to firmly establish the nature of the system. 

\begin{acknowledgments}
EBM and GGP acknowledge support from Penn State ACIS Instrument
Team Contract SV4-74018 (issued by the Chandra X-ray
Center, which is operated by the Smithsonian Astrophysical
Observatory for and on behalf of NASA under contract NAS8-
03060). This work is based on a Chandra ACIS Guaranteed Time Observation (GTO)
selected by the ACIS Instrument Principal Investigator, Gordon
P. Garmire, currently of the Huntingdon Institute for X-ray
Astronomy, LLC, which is under contract to the Smithsonian
Astrophysical Observatory via Contract SV2-82024.
\end{acknowledgments}



\software{\ciao\ v4.16 \citep{ciao},
          \texttt{HEASOFT} v6.33.2 \citep{heasoft},
          \acisextract\ v2023aug14 \citep{Broos2010},
          \sherpa\ v4.16 \citep{sherpa}
          }
          
\bibliographystyle{aasjournal}
\bibliography{v1460_gto}


\end{document}